\documentstyle[10pt]{article}
\begin{document}
\begin{titlepage}
\thispagestyle{empty}

\vspace*{2cm}

\begin{center}
{\Large \bf Two--nucleon emission in the longitudinal response}

\vspace{1.5cm}
{\large Giampaolo Co'}

\vspace{.3cm}
{Dipartimento di Fisica, Universit\`a di Lecce \\  and 
I.N.F.N. sez. di Lecce, I-73100 Lecce, Italy}

\vspace{1.5cm}
{\large Antonio M. Lallena}

\vspace{.3cm}
{Departamento de F\'{\i}sica
Moderna, Universidad de Granada, \\
E-18071 Granada, Spain}

\end{center}

\vspace{2cm}
\begin{abstract}
The contribution of the two--nucleon  emission in the
longitudinal response for inclusive electron scattering 
reactions is studied. The model adopted to perform the
calculations is based upon Correlated Basis Function 
theory but it considers only first order terms in the 
correlation function. The proper normalization of the 
wave function is ensured by considering, in addition to 
the usually evaluated two--point diagrams, also the 
three--point diagrams. Results for the $^{12}$C nucleus 
in the quasi--elastic region are presented. 
\end{abstract}

\vskip 1.cm
PACS number(s): 21.60.-n, 24.10.Cn, 25.30.Fj
\end{titlepage}

\setcounter{page}{1}
Electromagnetically induced two--nucleon knockout reactions
are considered to be well suited to study short--range
correlations (SRC) in nuclei \cite{tne95}. The basic idea is that
the real or virtual photon interacts with a correlated pair of
nucleons which are emitted from the nucleus.
Though the study of this process has been
proposed long time ago \cite{got58},  only recently,
with the advent of the high--intensity monochromatic photon beams
and 100\%--duty cycle electron beams, the technical
difficulties in performing this kind of experiments with
adequate statistics have been overcome.

The simple picture presented above involves
one-body  electromagnetic operators and short--range
correlations only, however, other mechanisms contribute to the
two--nucleon emission, for example meson exchange current (MEC)
and final state interactions, and this complicates the analysis
of the experimental data. 
It is therefore necessary to deal with experimental
situations where the alternative  emission mechanisms can be
disentangled, or to find kinematical
conditions where the emission via SRC becomes the dominant
one. 

For these reasons it is important to avoid those
energy regions dominated by collective excitations of the
nucleus, such as the giant resonance region,
because in these regions the multi--nucleon emission is
mainly induced by the residual interaction via the
excitation of many particle--many hole  configurations
\cite{kam96}. There is also another reason, a more
pragmatical one, to avoid the kinematical regions with
relatively low excitation energy.  In these regions the
excitation energy is just above the two--nucleon emission
threshold, the phase space available for the two nucleons
emitted is quite small and, as a consequence, the cross
sections are rather small.

It is therefore mandatory to work, at least, at the excitation
energies where the quasi--elastic peak shows up. In
this region, however, the emission mechanism we want to study,
competes with the two--nucleon emission produced by MEC
\cite{ama94}. Since MEC are active predominantly in the transverse
response, there is the hope that the two-nucleon emission in the
longitudinal response would be dominated by SRC effects.

In this paper we present the results of a 
calculation of the two--nucleon emission contribution to
the inclusive (e,e') longitudinal response. This calculation
has  been done for the $^{12}$C nucleus.

The model we have developed to describe the process
is based upon the Correlated Basis Function (CBF) theory 
\cite{cla79}, but it considers only the terms up to a single
correlation line.

In CBF theory the many--body Schr\"odinger
equation is solved by means of the variational principle
within a subspace of wave functions of the type:
\begin{equation}
\label{corst}
| \Psi \rangle \, = \, F \, | \Phi \rangle \, , 
\end{equation}
where $| \Phi \rangle$ is a Slater determinant built up with 
a set of single particle wave functions properly chosen, and $F$
is  the correlation function.
The variational method with the ansatz of Eq.~(\ref{corst}) 
has been successfully used to describe few-body
systems  \cite{arr95}, light nuclei \cite{pud97} and infinite 
systems \cite{wir88}, nevertheless, its application to medium and
heavy nuclei is still at the beginning stages.  Recently,
promising attempts to extend CBF theory to these last nuclear
systems have been carried out with the help of the Fermi
Hypernetted Chain (FHNC) technology \cite{ari96}. 

In addition to the known difficulties related to a FHNC
calculations in finite nuclear systems, our presents a further
complication due to the fact that it is necessary to
extend the FHNC theory to the description of the nuclear excited
states, in the same spirit of what has been done in
Refs. \cite{fan87,fab97} for nuclear matter.

For these reasons we have developed a model which is an
extension of those models used some time ago to
calculate ground state density and momentum
distributions \cite{kha68}. These models consider only
those  terms of the cluster expansion containing a single
correlation line.  A test of the validity of these models
have been recently done comparing their results with those
obtained using the same input in a full FHNC calculation
\cite{co95,ari97}. The good agreement obtained in this
comparison gives us the hope that a truncation of the
cluster expansion to the terms with a single correlation
line could work also for the description of nuclear
transitions, at least for those induced by  the charge
operator.

The basic hypothesis of our work lies in the ansatz of 
Eq.~(\ref{corst}).
The correlation function  $F$ is extremely
complicated and it has the same operator structure of the
nucleon--nucleon interaction. Realistic CBF calculations
\cite{arr95}--\cite{wir88} show that the scalar term of the
correlation function greatly dominates on the other ones.
This does not mean, however, that the so--called
state dependent terms of the correlation can be neglected because
their effect is small. For example, the tensor
correlations, are extremely important in the
calculation of the binding energy \cite{pan79}, and
probably they play a crucial role in setting the magnitude of
the MEC \cite{fab97,ang96} in the quasi--elastic peak.
On the other hand, in this work we consider the charge operator,
which has only an isospin dependence, and we believe that for
this operator the effects of the states dependent terms of the
correlation should be small.
For this reason, and to simplify the calculations, we have
considered a purely scalar correlation function of the form:
\begin{equation} F \, = \, \prod_{i<j}^{A} \, f(r_{ij}) \, , 
\end{equation} 
where $r_{ij}$ is the distance between the particles $i$ and $j$. 

The response produced by a generic operator $U({\bf q})$ is:
\begin{equation}
\displaystyle
\label{response}
R({\bf q},\omega) \, = \, \sum_f \, 
\frac{\langle \Psi_i | U^{\dag} ({\bf q}) | \Psi_f \rangle \,
      \langle \Psi_f | U ({\bf q}) | \Psi_i \rangle}
{\langle \Psi_i | \Psi_i \rangle \, \langle \Psi_f | \Psi_f \rangle}
\, \delta(E_f-E_i-\omega) \, .
\end{equation}
Assuming the same correlations for both ground and excited
states, the above equation can be rewritten in terms of the
amplitude: 
\begin{equation}
\label{xieq}
\displaystyle
\xi_{if}({\bf q}) \, = \, 
\frac{\langle \Phi_f | F^{\dag} U ({\bf q}) F | \Phi_i \rangle}
{\langle \Phi_i | F^{\dag} F | \Phi_i \rangle} \,
\left[
\frac{\langle \Phi_i | F^{\dag} F | \Phi_i \rangle}
{\langle \Phi_f | F^{\dag} F | \Phi_f \rangle} \,
\right] ^{1/2} \, ,
\end{equation}

This is the basic quantity to be studied and it
corresponds to the ground state expectation value of the
operator $U ({\bf q})$ in the case the state $| \Phi_f \rangle$
becomes  the ground state $| \Phi_i \rangle$. For the charge
operator, which  is the one we consider in our calculations, the
quantity  $\xi_{if}({\bf q})$ satisfies the following property:
\begin{equation}
\label{xilim}
\displaystyle
\lim_{q\rightarrow 0} \, \lim_{i\rightarrow f} \, 
\xi_{if}({\bf q}) \, = \, \frac{Z}{A} \, .
\end{equation}

To evaluate $\xi_{if}({\bf q})$, instead of performing
the full cluster expansion as it has  been done for
infinite nuclear systems \cite{fan87} we 
consider only terms of the cluster expansion 
containing a single correlation line $h$ defined as: 
\begin{equation} h(r_{ij}) \, = \, f^2(r_{ij}) - 1 \, .
\end{equation}

Cutting an infinite series is always a delicate operation because
a wrong choice of the terms retained can produce equations which
do not conserve the properties of the system under
investigation, such as the number of particles.   
In constructing our model we have been guided by the rule that the
terms considered should provide an approximate amplitude
$\xi^1_{fi}$  satisfying the limit of Eq.~(\ref{xilim}).

This model can be used to calculate transitions leading to
final states with one or two particles in the continuum. In
this work we are interested in the two--nucleon emission, and for
this process our model produces 4 two--point diagrams and
12 three--point diagrams. 

The three point diagrams describe the situation where three particle
are correlated, in spite of the  fact that only one two-point
(dynamical) correlation is  present.
In these diagrams, in addition to the dynamical correlation, 
also a statistical correlation, generated by the 
antisymmetrization of the many-body wave function under the exchange
of  two particle is acting.

Considering the symmetry properties
of the correlation function, $h(r_{ij})=h(r_{ji})$, and the fact
that some of this diagrams are obtained exchanging
particle and hole lines, these 16 diagrams reduce to the 8
topologically distinguished diagrams shown in Fig.~1.
A more thorough description of the model will be provided in a
forthcoming publication.

The calculations we discuss in the following have been done 
for the $^{12}$C nucleus. This nucleus is relatively light, it 
has only four hole single particle states, and therefore
calculations are less time consuming than for heavier nuclei.
In addition we have thoroughly studied the quasi--elastic
response of this nucleus \cite{ama93,ama94} and this
experience gives us some insight in the details of the 
configuration space to be used. In this respect, all our
calculations have been done with the set of single particle wave
functions generated by a Woods--Saxon potential used in our
previous quasi--elastic peak calculations. It is worth to notice
that the same  set of single particle wave functions has been used
in the FHNC calculations of Ref. \cite{ari96}. 

In Ref. \cite{ama94} we have calculated the two--nucleon
emission in the transverse response induced by the MEC. In the
present calculation we have used the same angular coupling scheme,
this time applied to the longitudinal response and for
the charge transition operator modified with the corresponding 
correlation term.

In Fig.~2 we present  
the various types of scalar correlation functions used to
test the sensitivity of our results to the details of the
correlation. The full and the dashed lines represent the Gaussian 
and ACA Euler correlations used in the FHNC calculations of Ref.
\cite{ari96}. These correlations have been fixed by minimizing the
binding energy of $^{12}$C for the Afnan--Tang S3 semirealistic
nucleon--nucleon interaction and for the same set of
single particle wave functions adopted in the present work. 

The third correlation we have used (dotted line) corresponds to the
scalar part of the Nuclear Matter correlation determined in
the FHNC calculations of Ref. \cite {wir88}.  
In addition we have also considered the
OMY correlation \cite{ohm56}, represented by the dashed dotted
line, because it has been widely used in the literature.

As far as we know, the (e,e'2N) calculations performed up to
now \cite{giu91,ryc95} consider only two--point diagrams (the
A and B diagrams of Fig.~1). In these calculations the
normalization of the wave function is not conserved because the
limit of Eq. (\ref{xilim}) is not satisfied. A first aspect to be
investigated with our model is then 
the importance of the three--point diagrams necessary to
fulfill Eq. (\ref{xilim}) at the first order in the correlation
line. 

In Fig.~3 we show the results we have obtained for
the contribution of the two--nucleon emission to the 
inclusive longitudinal responses for three values of the
momentum transfer. These calculations have been
performed with the Gaussian correlation.  
The full lines show the results found
with all the diagrams, while the dashed lines have been
obtained considering only the two--point diagrams.

This figure shows that the contributions of the two--
and three--point diagrams sum up to each other.
We have obtained analogous results for all the correlations
functions considered. This result is, in principle, surprising,
since, from our previous experience in the calculation of density 
and momentum distributions \cite{co95,ari97}, we expected big
cancelations between  two-- and three--point contributions.

In reality the correlations play different roles in the two 
cases.
In the ground state the correlations effects remodel the 
shape of the mean-field charge distribution without changing the 
total charge. 
In this case, every two--point diagram is coupled to a 
three--point diagram of opposite sign, which, in the limit of 
Eq.~(\ref{xilim}), cancels exactly the contribution of the two--point
diagram (the  uncorrelated charge distribution is already correctly
normalised). In the response the three--point diagrams offer an 
additional mechanism of emitting two nucleons, enlarging the 
available phase space, and therefore their contribution to the 
response adds up to that of the two-point diagrams.

A second aspect we want to investigate is the sensitivity of the
results to the correlation chosen. In Fig.~4 we show 
the full responses (that is including two-- plus three--point diagrams) 
obtained with the various correlation functions. The same convention as
in Fig.~2 has been used for the different curves. 
One should notice that, in the figure, the 
responses obtained with the OMY correlation (dashed dotted lines) have
been divided by a factor 10. 
All the other responses are of the same order of magnitude.

The results obtained with the
Gaussian correlation are very similar to those obtained with
the Nuclear Matter one, as expected because 
of the large similitude between these correlations
one can observe in Fig.~2.

The results of Fig.~4 show high sensitivity to the details of the
correlation function. Our approach does not provide any
prescription to choose among the correlations we have used. On the
other hand, in CBF theory the correlation functions are chosen
together with the single particle wave functions in a way to minimize
the ground state energy of the system.
The lack of an internal criterion to link single particle wave
functions and correlation is a weak point of our approach.
We think this problem can be overcome by taking these inputs
from a microscopic calculation of the ground state
energy.                

In the calculations we have presented, this has been done, at
least partially, for the Gaussian and ACA Euler correlations
which, for this set of single particle wave functions, minimize
the binding energy of $^{12}$C when the S3 nucleon--nucleon
interaction is used. There is not link between single particle
wave functions and the OMY correlation, which produces responses
one order of magnitude bigger than the other ones. 
In this sense  the comparison of Fig.~4 is not fully correct, 
because we should have
compared results obtained with correlations and single
particle wave functions modified to minimize the
nuclear binding energy. 

Let's summarize the main messages of this report.
\begin {enumerate}
\item In order to get the proper normalization of the wave 
function in a model considering only terms up to the first order in
the correlation line it is necessary to include both two-- and
three--point diagrams. 
\item The contribution of the three-point diagrams adds
strength to the response, contrary to the case of the ground
state expectation values, where a strong cancelation between 
two-- and three--point diagrams is found.
\item A relation between single particle wave functions and
correlation functions is necessary to have physically
meaningful results. 
\end{enumerate}

Before concluding we would like to make some general remarks 
about our model. As we have said, we infer the validity of
our model from the fact that the results of the ground
state density and momentum distribution were quite similar
to those obtained with the FHNC calculation \cite{ari97}. 
We do not claim that models considering only first order terms
in the correlation can be blindly applied to any operator. We
believe that the good results obtained for the ground state
expectation values of the charge distribution are related to
the peculiar characteristics of this operator. Because of this, we
feel quite confident of our model devised for the calculation of
responses, but we think that  a comparison with FHNC responses is a
necessary test. Work in this direction is in progress.

\newpage

\noindent
\section*{Figure Captions}

\vskip 0.5 cm

\noindent
FIGURE 1 \\
Diagrams considered in our model. The dotted lines represent the
correlation function $h$. The full oriented lines represent
particle and hole wave functions. We have indicated with $p1$, $p2$,
$h1$ and $h2$ the wave functions of the two particles and two hole
states surviving asymptotically, and with $\alpha$ a generic hole
wave function different from $h1$ or $h2$. The black circle
indicates an integration point while the black square indicate the
integration point where the electromagnetic operator $U(q)$, the
charge operator in our case, is acting. In addition to these diagrams 
we consider also those obtained by
exchanging the pairs ($p1,h1$) with the pairs ($p2,h2$), as well as
those three--points diagrams where the two points linked by the
correlation function are exchanged.

\vskip 0.5 cm

\noindent
FIGURE 2 \\
Correlation functions used in our calculations. The full and
dashed lines are the Gaussian and the ACA Euler correlations of
Ref.~\cite{ari96}. The dotted line is the scalar part of the 
Nuclear Matter correlation function of Ref.~\cite{wir88}. 
The dashed--dotted line represent the OMY correlation 
function~\cite{ohm56}.

\vskip 0.5 cm

\noindent
FIGURE 3 \\
Inclusive longitudinal response functions for the emission of
two nucleons calculated for three different values of the momentum
transfer. The calculation has been performed with the Gaussian
correlation function. The full lines show the result obtained
considering all the diagrams presented in Fig.~1, while the dashed
lines have been obtained only with the two--point diagrams
(diagrams A and B in Fig.~1).

\vskip 0.5 cm

\noindent
FIGURE 4 \\
Inclusive longitudinal response functions for the emission of
two nucleons calculated for three different values of the momentum
transfer. The various lines represent the results obtained with the
correlations of Fig.~1. The full
lines have been obtained with the Gaussian correlation, 
the dashed ones with the Euler ACA, the dotted ones with the
Nuclear Matter correlation, and the dashed--dotted ones with the
OMY correlation. Note that the OMY dashed--dotted curves have been
multiplied by a 0.1 factor.

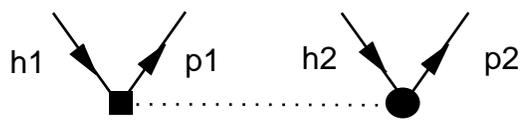
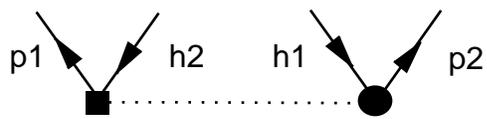
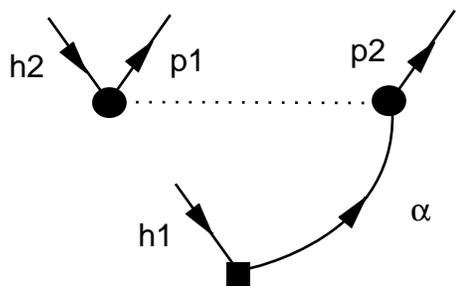
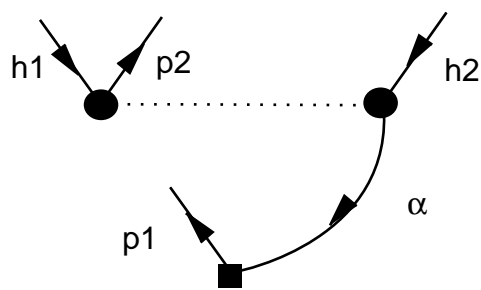
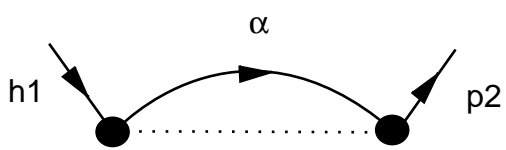
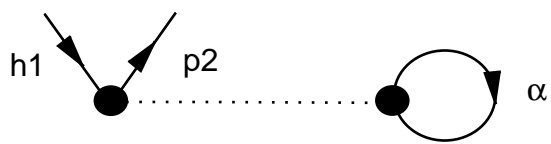
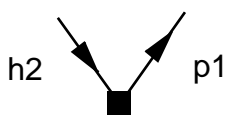
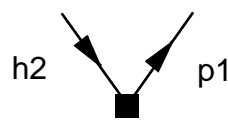
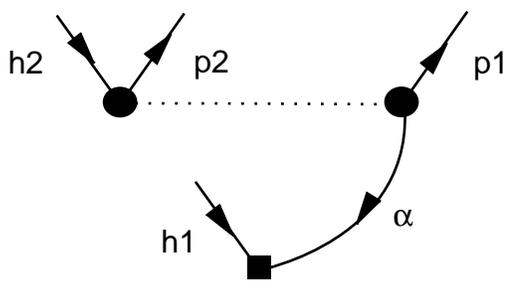
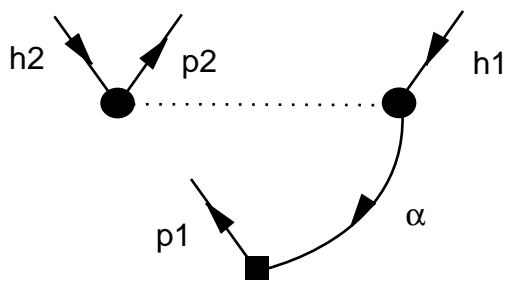
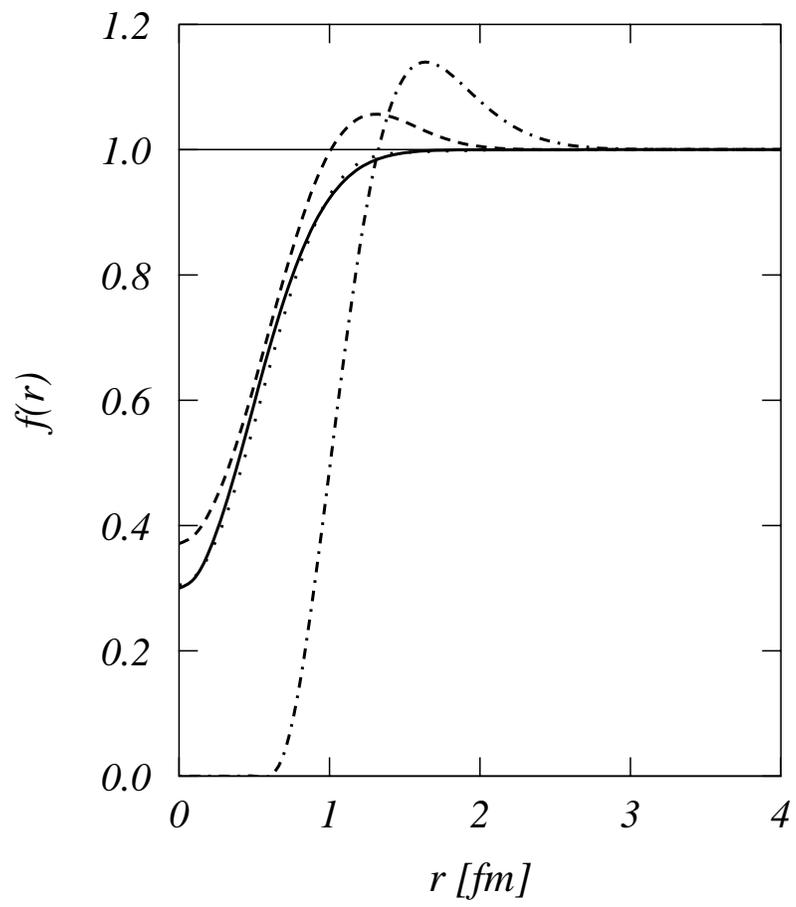
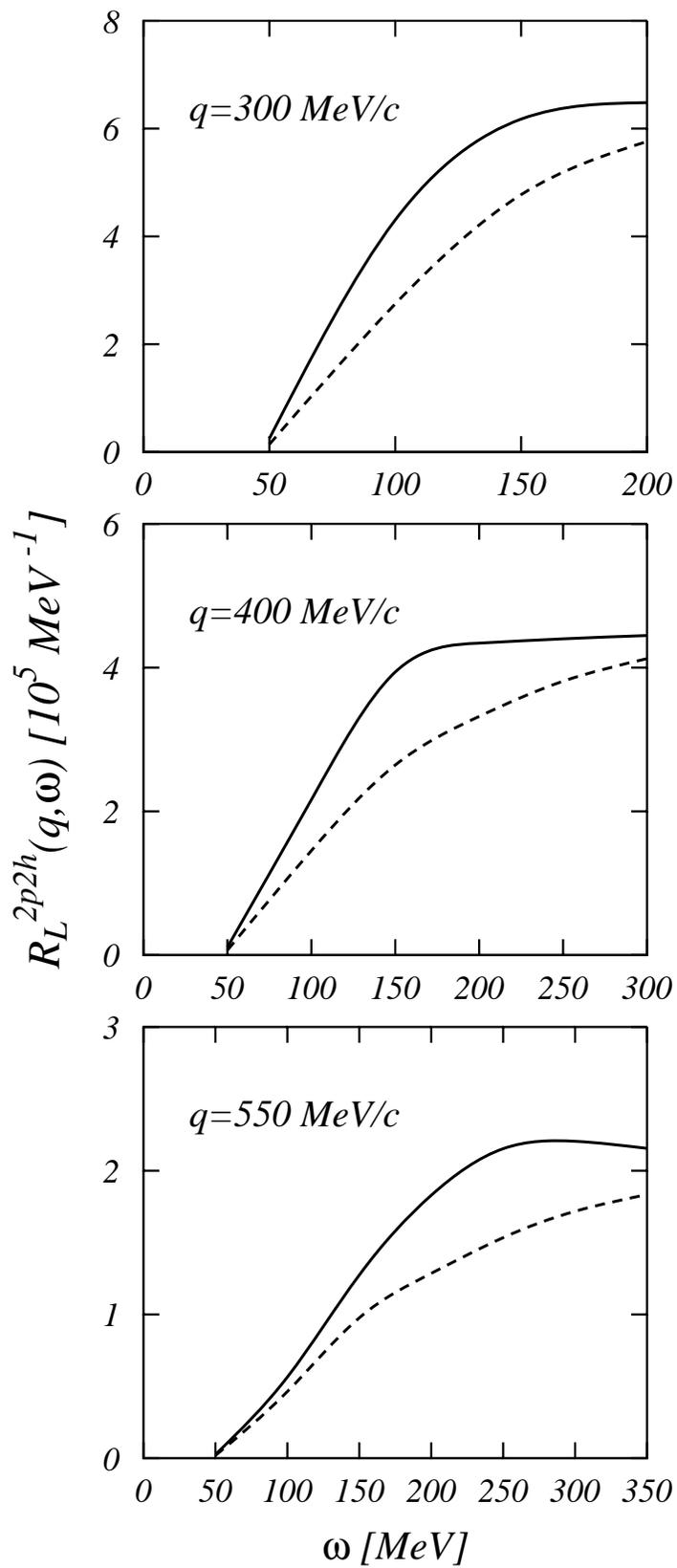
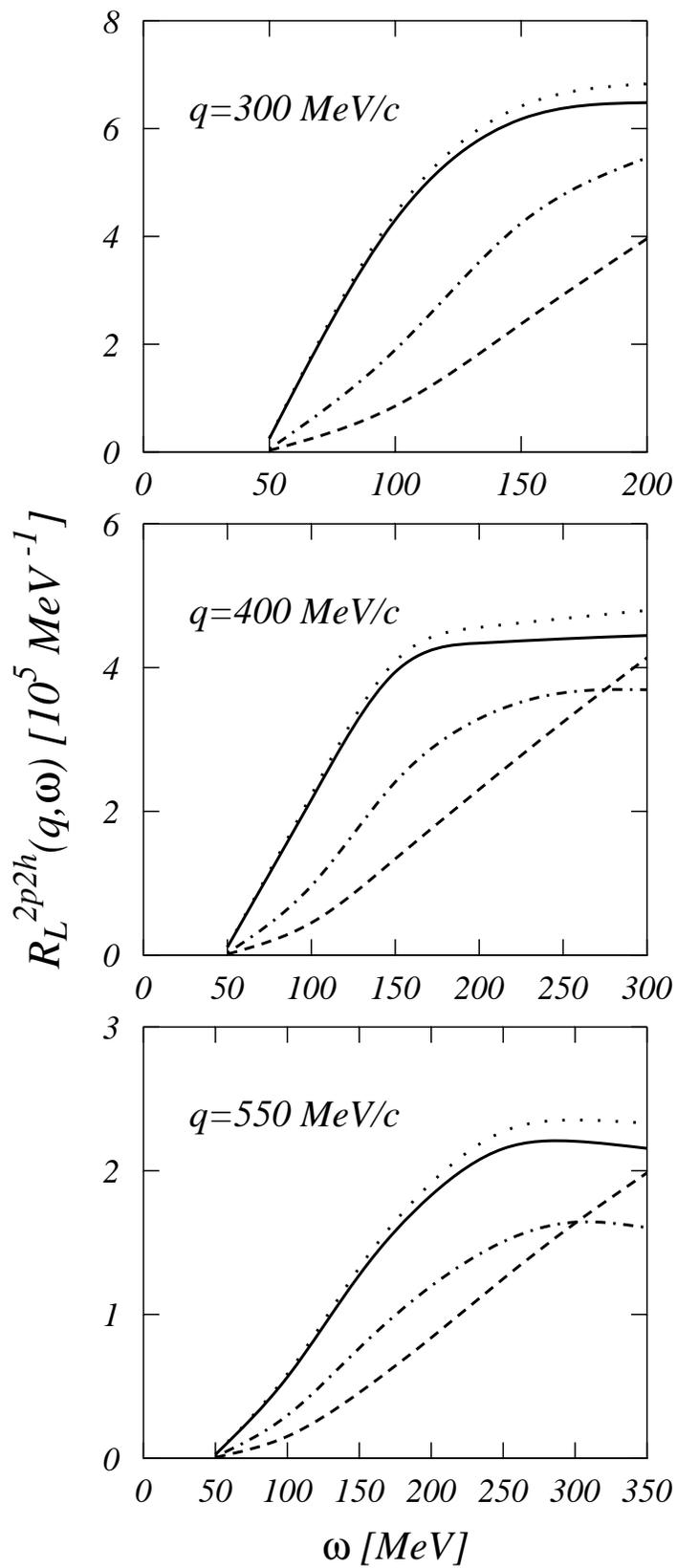
\end{document}